\documentclass[11pt]{article} %<-- this is for arxiv.  Default is 12

\usepackage{verbatim}
\usepackage{times}
\usepackage{graphicx} 
\usepackage{chngcntr}

\usepackage{booktabs}
\usepackage{amsmath}

\usepackage{url}

\usepackage{xcolor} %colors,yay

\usepackage{setspace}
\usepackage{comment}

\newenvironment{sciabstract}{%
\begin{quote} \bf}
{\end{quote}}

\title{Nuclear Disarmament Verification via Resonant Phenomena }

\author
{Jake J. Hecla, Areg Danagoulian$^{\ast}$\\
\\
\normalsize{Department of Nuclear Science and Engineering, Massachusetts Institute of Technology,}\\
\normalsize{77 Massachusetts Avenue, Cambridge, MA 02139, USA}\\
\\
\normalsize{$^\ast$To whom correspondence should be addressed; E-mail:  aregjan@mit.edu.}
}

\date{}

\begin{document}


%\baselineskip24pt
%single space
\baselineskip12pt  %<-- this is for arxiv.  For others revert to 24pt


\maketitle


\begin{sciabstract}

%The abundance of nuclear weapons is one of the biggest existential threats to human civilization.  Ambitious nuclear disarmament treaties are necessary to reduce this danger, along with the technologies necessary for verifying adherence to these treaties.  This includes verifying the authenticity of the nuclear warheads undergoing dismantlement before counting them towards a treaty partner's obligations. 

Nuclear disarmament treaties are not sufficient in and of themselves to neutralize the existential threat of the nuclear weapons. Technologies are necessary for verifying the authenticity of the nuclear warheads undergoing dismantlement before counting them towards a treaty partner’s obligation. This work presents a novel concept that leverages isotope-specific nuclear resonance phenomena to authenticate a warhead's fissile components by comparing them to a previously authenticated template. All information is encrypted in the physical domain in a manner that amounts to a physical zero-knowledge proof system. Using Monte Carlo simulations, the system is shown to reveal no isotopic or geometric information about the weapon, while readily detecting hoaxing attempts.  This nuclear technique can dramatically increase the reach and trustworthiness of future nuclear disarmament treaties.   

%Old
%The abundance of nuclear weapons is one of the biggest existential threats to human civilization, and ambitious, aggressive nuclear arms reduction treaties are necessary for reducing this danger. To enable these treaties, however, technologies are necessary to verify the authenticity of the nuclear warheads undergoing dismantlement before counting them towards a treaty partner's obligations. We present a novel concept which leverages isotope-specific resonance phenomena to authenticate a warhead's fissile components by comparing them to a previously authenticated template. All information is encrypted in physical domain in a manner which amounts to a physical zero-knowledge proof system. Using Monte Carlo simulations, the system is shown to reveal no isotopic or geometric information about the weapon, while readily catching cheating attempts.

%  This document presents a number of hints about how to set up your
%  {\it Science\/} paper in \LaTeX\ .  We provide a template file,
%  \texttt{scifile.tex}, that you can use to set up the \LaTeX\ source
%  for your article.  An example of the style is the special
%  \texttt{\{sciabstract\}} environment used to set up the abstract you
%  see here.
\end{sciabstract}

% In setting up this template for *Science* papers, we've used both
% the \section* command and the \paragraph* command for topical
% divisions.  Which you use will of course depend on the type of paper
% you're writing.  Review Articles tend to have displayed headings, for
% which \section* is more appropriate; Research Articles, when they have
% formal topical divisions at all, tend to signal them with bold text
% that runs into the paragraph, for which \paragraph* is the right
% choice.  Either way, use the asterisk (*) modifier, as shown, to

\paragraph*{Summary:}  The resonant responses triggered by epithermal neutrons in nuclei can be leveraged for a zero-knowledge verification of nuclear warheads.

\section*{Introduction}

The abundance of nuclear weapons could be the biggest existential threat to human civilization.  Currently Russia and the US own more than 90\% of all nuclear weapons. It is estimated that as many as 14 000 units are part of their arsenals of retired and stockpiled weapons, and an additional 3 700 units make up the deployed arsenals~\cite{Kristensen_US,Kristensen_Russia}.  Such large numbers expose the world to the danger of catastrophic devastation in case of an intentional or accidental nuclear war, and also raise the danger of nuclear terrorism and nuclear proliferation due to possible theft or loss of nuclear weapons. So far most treaties, such as the Strategic Arms Reduction Treaty (New START) and the Intermediate-Range Nuclear Forces (INF) treaty, stipulate cross-verification of the dismantlement of delivery systems -- such as bomber aircraft and cruise missiles.  The delivery methods are a reliable proxy of strike capability in a nuclear war scenario.  However, a reduction effort that is limited only to the verification of delivery methods leaves behind the problem of large stockpiles of surplus nuclear warheads.  Thus disarmament treaties that target the stockpiles themselves are indispensable for reducing this combined danger. To enable such treaties, however, new technologies are necessary to achieve treaty verification while protecting the secrets of the treaty participants.   This means verifying the authenticity of nuclear weapons -- before their destruction is counted towards a treaty participant’s obligations -- without revealing any classified information.  Such a technique will be a highly powerful tool in enacting far-reaching disarmament treaties.

But how does one verify that an object is a weapon without inspecting its interior?  This apparent paradox has puzzled policy makers and researchers alike for the last few decades, with no clear solutions adopted.  Past US-Russia lab-to-lab collaboration included the research and development of information barriers (IB)~\cite{ref:Hecker}.  These are devices that rely on software and electronics to analyze data from radiation detectors and compare the resulting signal against a set of attributes in a so-called attribute verification scheme~\cite{fuller}. The attributes can be the plutonium mass and enrichment, the presence of explosives, etc.  Most importantly, these attributes have to be quite broad, to prevent the release of classified information.  This in its turn makes them insensitive to a variety of hoaxing scenarios.  The other major difficulty of this paradigm is that it shifts the problem of verification to software and electronics, whose components themselves will need to undergo verification and validation for the possible presence of spyware, back-door exploits, and other hidden functionality.  If present, these can either leak secret information to the inspectors, or clear fake warheads.  To overcome this, an alternative approach called template verification has been proposed where all candidate warheads and/or their components are compared to those from a previously authenticated template. The authenticated template itself can be selected based on situational context:  a random warhead acquired from a deployed ICBM during a surprise visit by the inspection crew can be expected to be real, since a country's nuclear deterrence hinges on having real warheads on their ICBM.  To strengthen the confidence in its authenticity, multiple warheads can be removed from multiple ICBM and later compared to each other using the verification protocol described in this work.

While significant first steps were taken in template verification research, much needs to be done in insuring that the verification protocol is both hoax-resistant as well as information secure.  A template verification method based on the non-resonant transmission of fast neutrons was developed by researchers at Princeton~\cite{ref:alex}. Independently,  an isotope-sensitive physical cryptography system was proposed by researchers, including one of the authors of this paper, at MIT's Laboratory of Nuclear Security and Policy~\cite{PNAS,vavrek2016thesis,ref:ANS}.  Both approaches have advantages and disadvantages.  The Princeton concept has strong information security in the form of a zero-knowledge (ZK) proof.  It relies primarily on the non-resonant scattering of fast neutrons, a process which is almost identical for most actinides, making it prone to isotopic hoaxes, e.g.  via replacement of the weapons grade plutonium (WGPu) pit{\footnote{In this work the pit refers to the hollow plutonium sphere at the center of a fission nuclear weapon.}} with easily available depleted uranium (DU) or reactor grade plutonium (RGPu).  The previous MIT system, relying on isotope-specific Nuclear Resonance Fluorescence (NRF) signatures offers strong hoax resistance.  However it is not fully zero-knowledge, and thus needs to undergo thorough checks for information security.  The new methodology proposed in this work combines the strengths of the Princeton concept (ZK proof) with the strengths of the MIT concept (isotopic sensitivity), while avoiding their weaknesses.  Such an outcome can have a strong impact on future arms reduction treaties. 

The new technique  uses resonance phenomena to achieve isotope-specific data signatures, which can be used to obtain the necessary "fingerprint" of the object.  This is achieved by exploiting nuclear resonances in actinides when interacting with epithermal neutrons in the 1-10 eV range.  Unlike fast neutrons, which do not have this isotope-specificity for high Z nuclei and thus cannot enable a crucial resistance to isotopic hoaxes, the epithermal neutron transmission signal can be made highly specific and sensitive to the presence and abundance of individual isotopes.  These include $^{235}$U and $^{239}$Pu in highly enriched uranium (HEU) and WGPu~\cite{ref:chichester2012assessing}. Also unlike NRF, the resonant absorption of epithermal neutrons in the beam can be observed directly with very high resolution (less than eV).   This can be done by using time-of-flight (TOF) techniques.  These characteristics allows for direct measurements of resonant absorption, thus enabling zero-knowledge implementations otherwise impossible with NRF.

\section*{Protocol}

The key to any verification procedure is a protocol that can guarantee that no treaty accountable item (TAI) undergoing verification is secretly modified or replaced with another object.  The general steps of the protocol are as follows:

\begin{enumerate}
\item The inspection party (inspectors) makes an unannounced visit to an ICBM site and randomly chooses a warhead from one of the missiles.  The warhead enters the joint custody of the inspectors and the host country (hosts), and can be treated as the authentic template to which all future candidate warheads undergoing dismantlement and disposition will be compared.
\item  The template is transported under the joint custody of the hosts and inspectors to the site where the candidate warheads (henceforth referred to as candidates) will undergo dismantlement, verification, and disposition.
\item  Both the template and the candidate undergo dismantlement by the hosts in an environment that cannot be observed directly by the inspectors, but from where no new objects can be introduced to or removed.  This could be done by curtaining off an area in the middle of a hall. The fissile component of the weapon, also known as the pit, is extracted.  The pits are placed in marked, opaque boxes.  The remaining components are placed in a different, unmarked box.
\item  The curtains are removed and the dismantlement area is made accessible to the inspectors. The task of the inspection is to verify that the template and candidate pits in the marked boxes are geometrically and isotopicaly identical.  The inspectors use a Geiger counter to verify that the non-fissile, unmarked boxes do not contain any radioactive materials, thus confirming that the marked boxes indeed contain the pits.
\item The two marked boxes undergo epithermal transmission analysis.  The 2D radiographs and the spectral signatures are compared in a statistical test.  An agreement confirms that the candidate is identical to the template and thus can be treated as authentic.  A disagreement indicates a hoaxing attempt. 
\end{enumerate}
The last step is key to the whole verification process, and is the focus of this work.  The transmission measurement produces an effective image in a 2D pixel array made of scintillators that are sensitive to epithermal neutrons.  Furthermore, by choosing a detector with a fast response time and knowing the time when the neutrons are produced, a TOF technique can allow an explicit determination of neutron energies.

\section*{The Method}

\begin{figure}[ht]
    \centering
    \includegraphics[width=0.5\textwidth]{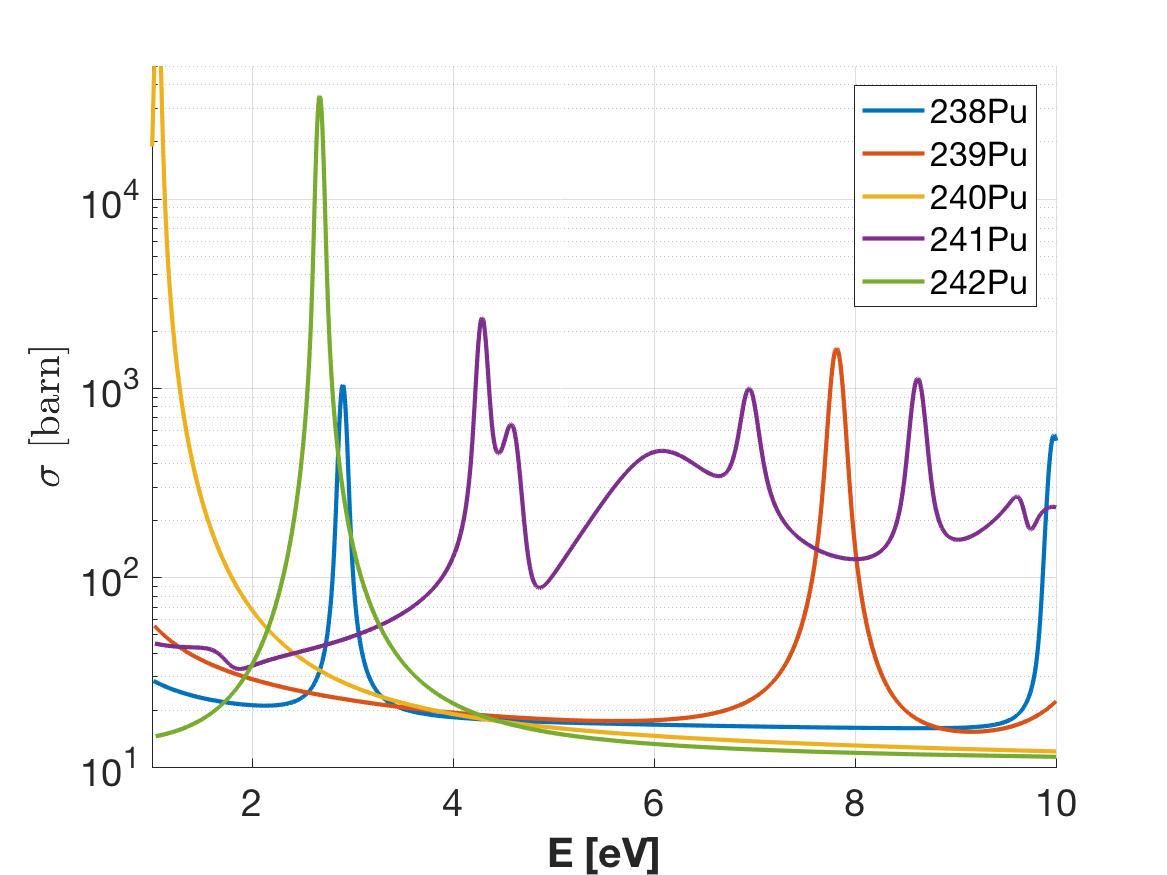}
    \noindent\caption{Total interaction cross sections for epithermal neutrons for various plutonium isotopes, showing very clear differences.  WGPu is almost entirely made out of $^{239}$Pu and $^{240}$Pu, while RGPu contains significant contributions from $^{240,241,242,238}$Pu.   Evaluated data taken from the JEFF-3.2 database~\cite{ref:koning}.}
    \label{fig:epithermal}
\end{figure}

The epithermal range refers to the neutron energy domain encompassed between the thermal energies of $\sim$40 meV and fast neutron energies of $\sim$100 keV.  The energies of interest for this study are those of $1 \le E \le 10$ eV.  While the neutron interactions in the thermal regime are described by monotonic changes in cross sections, in the epithermal range the neutrons can trigger various resonant responses in uranium and plutonium.  These are typically (n,fission) or (n,$\gamma$) reactions, resulting in the loss of the original neutron.  A plot of total interaction cross sections in the epithermal range for five isotopes of interest can be seen in Fig.~\ref{fig:epithermal}.  For a radiographic configuration these interactions selectively remove the original neutrons of resonant energies from the transmitted beam and give rise to an absorption spectrum, resulting in unique sets of $\sim$0.3 eV wide notches specific to each isotope.  While the resonances are the most prominent features of the cross section, the continuum between the resonances also encodes information about the isotopic compositions of the target. These combined absorption features yield a unique "fingerprint" of a particular configuration of isotopics, geometry, and density distribution.  This feature has been used in the past for non-destructive assay of nuclear fuel~\cite{ref:chichester2012assessing}.

\begin{figure}[ht]
    \centering
    \includegraphics[]{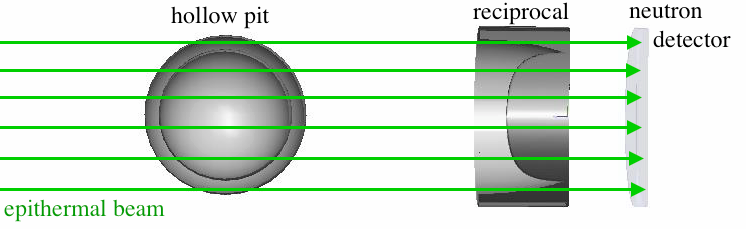}
    \noindent\caption{A diagram of the pit and its reciprocal mask, aligned along the axis of the interrogating beam.  The combined transmission image will be identical to that of a flat plate with a thickness equal to the external thickness of the mask. }
    \label{fig:reciprocal}
\end{figure}

The key component of a nuclear weapon is its pit - the hollow plutonium sphere at the center of the assembly. This work focuses on the verification of the authenticity of the pit.  A direct transmission imaging of the pit would reveal many of its secret parameters.  Thus, an additional physical cryptographic barrier is necessary to achieve a ZK test.  The barrier needs to be constructed such as to also allow for pit-to-pit comparisons that can detect any significant differences.  These two simultaneous goals can be accomplished by the mass reciprocal mask of the pit.  The reciprocal mask has a shape such that all the rays of epithermal neutrons in the beam transit the same combined areal density.  A simple example of a reciprocal mask of a hollow shell could be a cube with the geometry of the said hollow shell subtracted.  Thus, the aligned combination of the pit and the reciprocal will result in uniform areal density, producing an image in the detection plane which is consistent with that of a flat object.  A more optimal reciprocal for a hollow shell of internal and external radii $r_1$ and $r_0$ can be defined via its thickness along the beam axis $d=D-2(\sqrt{r_0^2-y^2}-\sqrt{r_1^2-y^2})$ for $y<r_1$ and $d=D-2\sqrt{r_0^2-y^2}$ for $r_1 \le y \le r_0$, where $D$ is the combined thickness observed by all particles in the beam, and $y$ is the vertical coordinate.  A combination of the pit and the reciprocal is illustrated in Fig.~\ref{fig:reciprocal}.  In order to keep the shape of the reciprocal secret, the hosts can place it inside a n opaque box.  It should be noted that an assembly with a uniform areal density by no means implies that the resulting image will be flat as well.  Secondary processes, e.g. neutron scattering, can distort the image and introduce some dependencies that contain information about geometric structures.  For this reason detailed simulations are necessary to validate the concept and demonstrate information security.  In the next sections we use Monte Carlo simulations to show that the transmission image produced for this configuration is identical to that of a uniform plate.  This outcome guarantees that no geometric information is revealed by the transmission analysis.  Even if the inspectors are capable of determining the incident flux in the neutron beam they will at most gain knowledge of an upper limit on the amount of total mass. The hosts can modify the mask to make that knowledge of no value:  for example, the inspectors might determine that the mass of WGPu in the pit is less than 10 kg - this knowledge is useless, as the critical mass for a WGPu pit is approximately 6 kg. 

\subsubsection*{Isotopic Hoax Resistance}

As the inspectors perform the radiographic interrogation of the pit, they need to ascertain that the isotopics and effective densities of the template and the candidate pits are identical.  At a given pit orientation, this can be performed by comparing the epithermal spectra of the transmitted beam from the template and candidate measurements.  The energy information can be acquired from the timing of the arrival of the neutrons, via the previously described TOF method to very high precision with either boron doped microchannel plate detectors, or $^6$Li based LiCaAlF$_6$  scintillators~\cite{ref:tremsin,ref:YAMAZAKI}.

To study the system's sensitivity to hoaxing, we consider a scenario where the WGPu pit has been replaced with a RGPu pit.  MC simulations are performed for the configuration described in Fig.~\ref{fig:reciprocal}.  The simulations were performed using the MCNP5 package, which has a fully validated neutron physics module~\cite{ref:mcnp}. The neutron energies were uniformly sampled in the $0 \le E \le 10$~eV range.  For this study the isotopic mass concentrations for RGPu was 40.8\% $^{239}$Pu.  For WGPu the fractions are 93\% $^{239}$Pu, the remainder as $^{240}$Pu and trace amounts of the other isotopes.  See supplementary materials for detailed listing of the isotopic concentration. For both the template and the hoax RGPu pit the inner diameter was 6.27 cm and the outer diameter was 6.7 cm - based on public domain estimates of a Soviet tactical thermonuclear warhead pit geometries~\cite{ref:fetter1990gamma}.  The reciprocal was modeled in three dimensions along the previously described formula.  The combined pit-reciprocal thickness was 5 cm.

\begin{figure}[ht]
    \centering
    \includegraphics[width=0.9\textwidth]{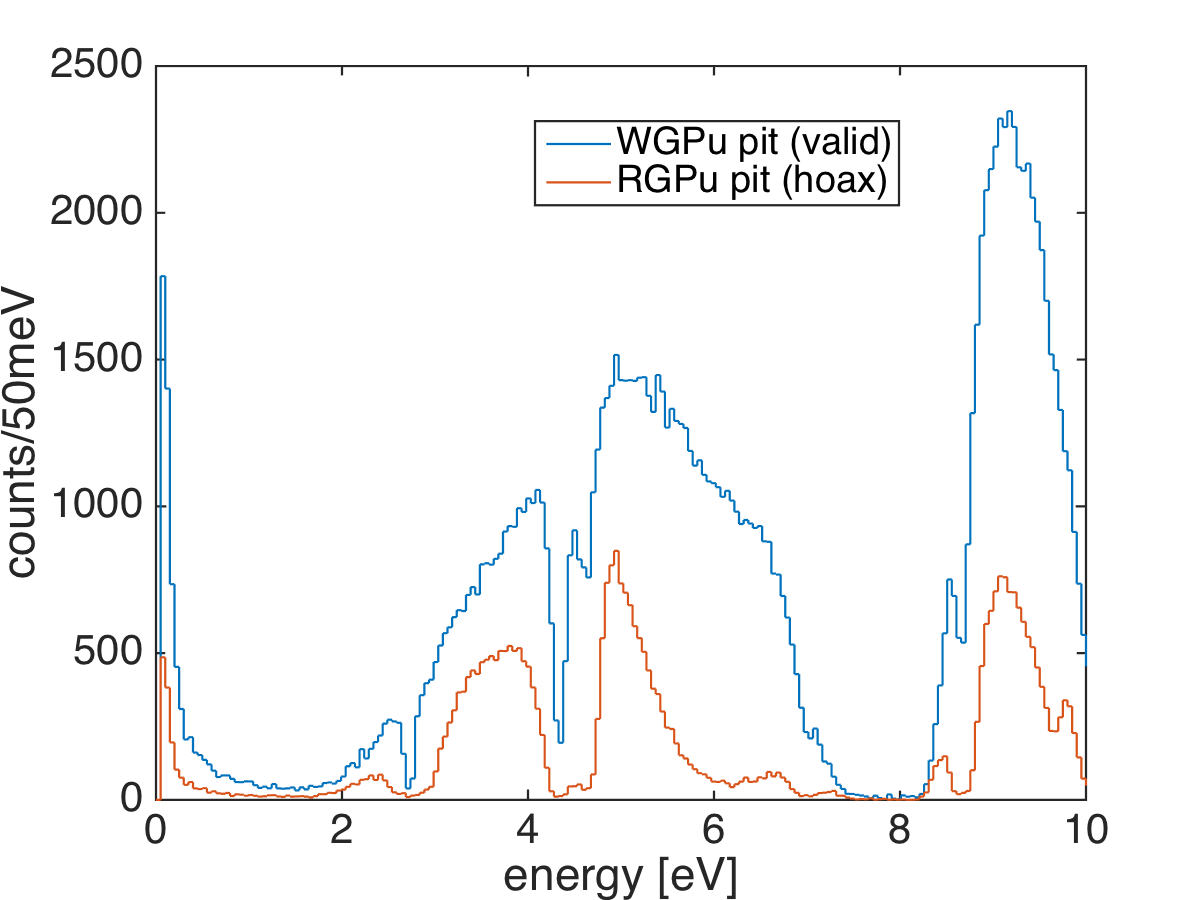}
    \noindent \caption{Simulations of transmitted epithermal flux spectra in an idealized detector for a valid WPGu-reciprocal and hoax RGPu-reciprocal configurations.} \label{fig:wgpu_vs_rgpu}
\end{figure}

Fig.~\ref{fig:wgpu_vs_rgpu} shows the results of the Monte Carlo simulation of an idealized detector exposed to the transmitted flux through a combination or a reciprocal mask and a WGPu pit, and a similar configuration where WGPu has been replaced by RGPu.  The large spectral discrepancies are clear.  The difference is primarily caused by the abundance of $^{242}$Pu in the RGPu, produced as a results of neutron capture in the reactor core.  This manifests itself in deeper absorption lines at 4.2 and 8.5 eV, as well as generally increased absorption in the 5-8 eV range.  For this simulation $20.7 \times 10^6$ neutrons were uniformly sampled in the 0-10 eV range, and the energy of the output was histogramed over 202 bins. For two given spectra, the chi-square test can be applied to reject or accept the null hypothesis, i.e. the hypothesis that the fluctuations are merely statistical and normally distributed.  For this result the value is $\chi^2=67177$, translating to a confidence for rejecting the null hypothesis of essentially exactly $p=1.0$ and showing conclusively that a hoaxing attempt is underway.  Furthermore, it is possible to determine the minimum number of incident epithermal neutrons necessary for achieving a confidence of $p=1-2.9\times10^{-7}$, which corresponds to the standard $5\sigma$ test.  That number is $n=1 \times 10^5$ neutrons.  For comparison, the MIT research reactor has been used to produce $10^{10}$ n/s/cm$^2$ epithermal neutrons in the $[1$ eV$,10$ keV$]$ range~\cite{ref:harling}. For this beam it corresponds to $\sim10^{9}$ n/s in the $[1,10]$ eV range. Thus a measurement requiring $10^5$ neutrons would require fractions of a second.  Alternative sources of epithermal neutrons are also possible, e.g. via nuclear reactions triggered by compact proton accelerators~\cite{ref:herrera2015new}.  Such sources would produce enough epithermal neutrons to achieve necessary measurements in about 12 seconds.  See supplementary materials for additional discussion and detailed calculations.

\subsubsection*{Geometric Hoax Resistance}\label{sec:geom_hoax_resistance}

\begin{figure}[ht]
    \centering
    \includegraphics[width=0.3\textwidth]{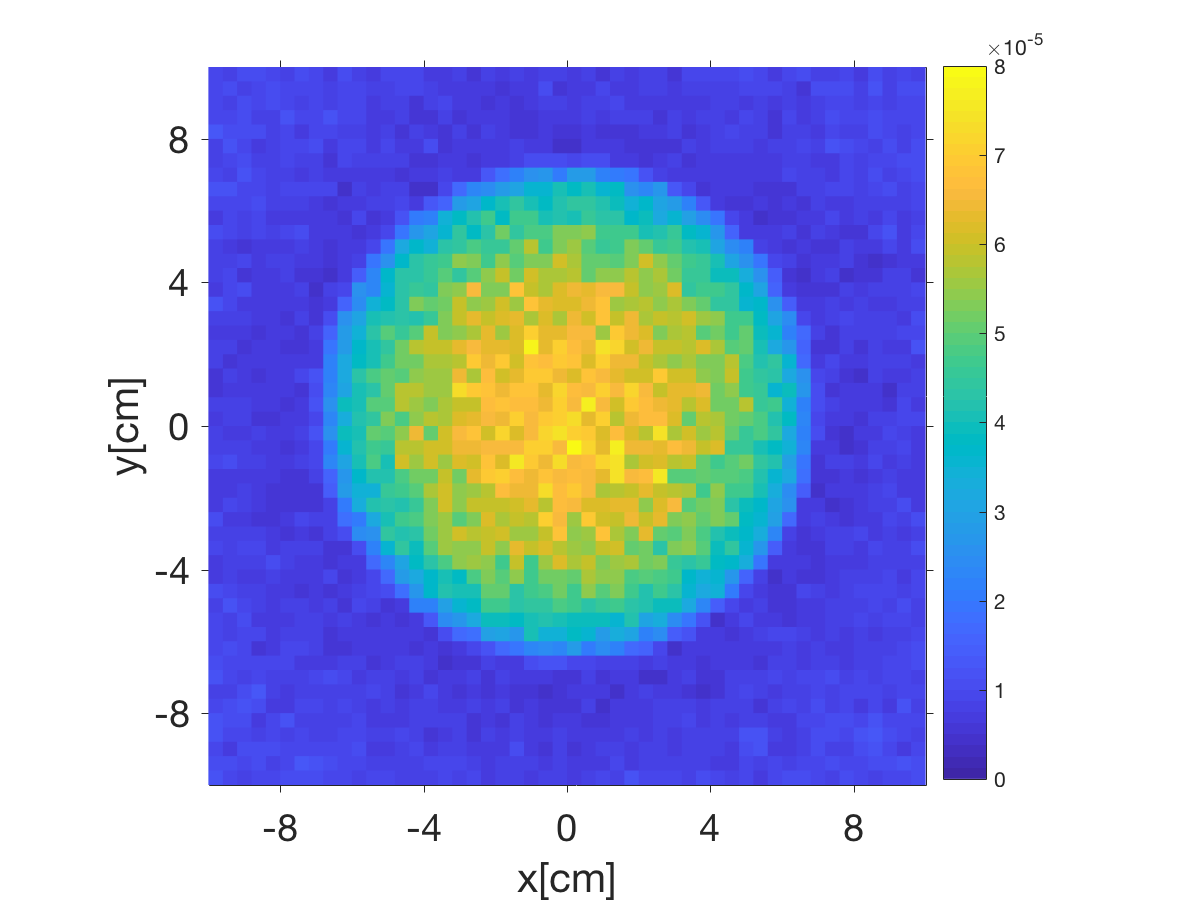}
    \includegraphics[width=0.3\textwidth]{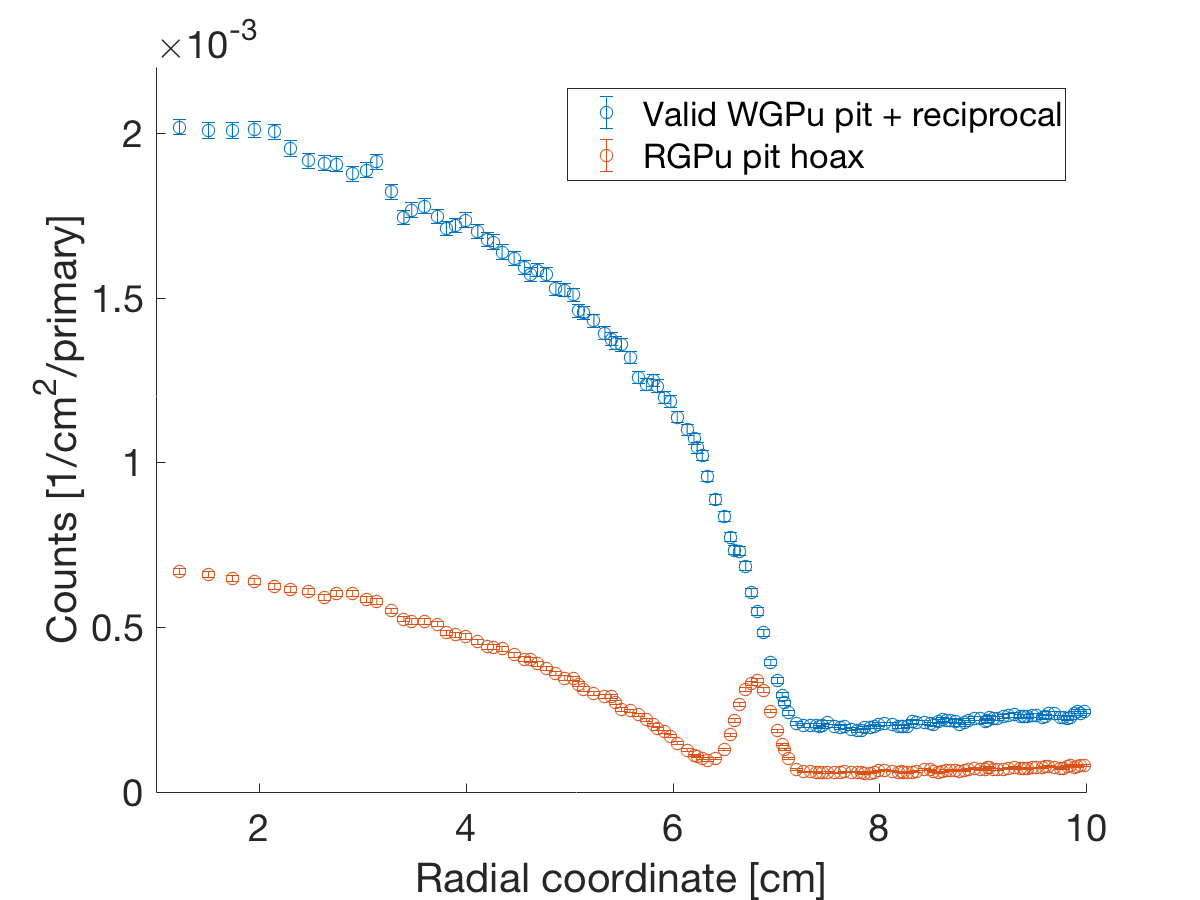}
    \includegraphics[width=0.3\textwidth]{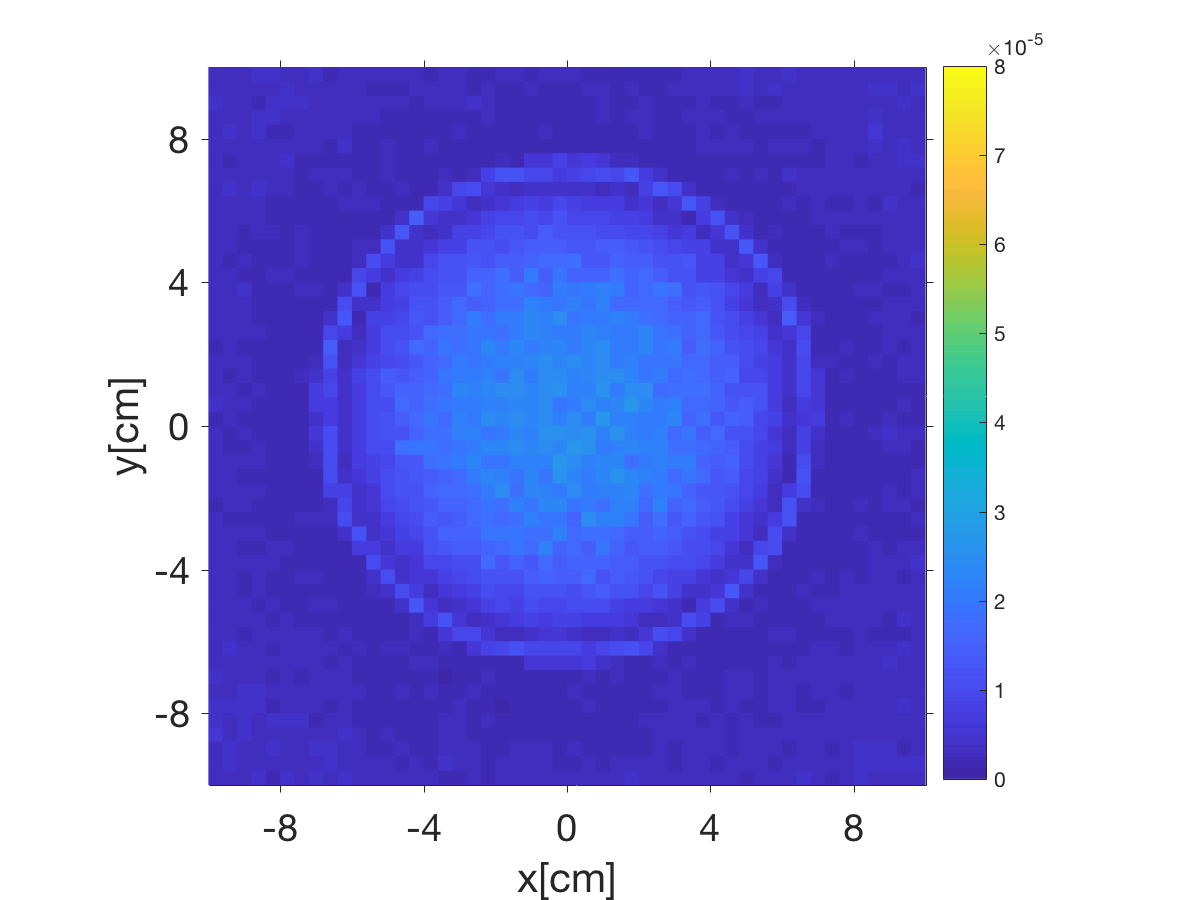}

    \noindent\caption{Epithermal neutron images in the detector for the WGPu-reciprocal (left) and for RGPu-reciprocal  hoax scenario (right),, and the radial distribution of the counts for both scenarios (center). 
    showing a clear discrepancy. The non-flat image of the WGPu configuration is caused by the in-scatter of the epithermal neutrons. 
    %The z-axis is the number of counts per incident neutron.
    }
    \label{fig:wgpu_vs_rgpu_2d}
\end{figure}

Geometric hoax resistance implies the use of 2-dimensional transmission imaging to identify any significant geometric and/or isotopic differences between the template and the candidate, as a way of detecting cheating attempts.  The detector described in Fig.~\ref{fig:reciprocal} and modeled in the MC simulation, whose spectral output is shown in Fig.~\ref{fig:wgpu_vs_rgpu}, can be pixelated to produce imaging data.  Fig.~\ref{fig:wgpu_vs_rgpu_2d} shows the results of simulated epithermal transmission images for a WGPu template and RGPu hoax candidate scenarios.  The 2D images show a clear difference, indicating that the candidate is a hoax and thus confirming the isotopic analysis described earlier.  The 1D projection of the two images' radial distributions further show the extent of the discrepancy. It should be noted that the non-flat image of the WGPu configuration is caused by the in-scatter of the epithermal neutrons, and doesn't reveal any information about the object, as will be shown in the section on geometric  information security.  The imaging analysis described here focuses on a hoax scenario of modified isotopics.  However any changes in pit diameter(s) and/or density  will result in non-uniformity in areal density as observed by the incident epithermal beam and will thus cause imaging inconsistencies similar to the ones described in Fig.~\ref{fig:wgpu_vs_rgpu_2d}.

While the above discussion is indicative of the strength of the technique in detecting hoaxes, a proof of uniqueness is necessary. This means showing that for a given reciprocal a given image can be produced only by a unique object.  This is not the case for a single projection, since transmission is only sensitive to the line integral along the beam axis.  To exclude the possibility of geometric hoaxes with identical line integrals but consisting of a different three dimensional manifold, multiple projections along the x, y, and z axes may be necessary.  For a spherically symmetric pit similar to the one treated in this work this could be achieved by multiple measurements at random angles.  For a spherically non-symmetric objects three projections along the x,y, and z axes may be sufficient  - each requiring a reciprocal built for that projection. A more rigorous proof of uniqueness is part of future work, and could be based on the methodology of K-transforms as described in Ref.~\cite{PNAS}.

%% Include if the reviewers have questions...
%\section*{Information Security}
%The protocol needs to be designed such as not to reveal any sensitive information to the inspectors, as they perform the verification procedure.  This means that the inspectors can learn nothing more about the pit geometry and/or isotopic composition than they already knew.  The reciprocal mask is the key to achieving that goal.  It is to be built and provided by the hosts in an optically opaque box.  The inspectors can ascertain that the reciprocal mask is not moved between the measurements, however they have no knowledge of its composition and shape.  The hosts may choose to optimize the reciprocal for some hoax candidate - yet they cannot do this simultaneously for a hoax candidate and an authentic template. 

\subsection*{Geometric Information Security}

Geometric information security refers to the notion that the inspectors will learn nothing about the pit geometry beyond what they already know.  We assume that for the proof system to be zero-knowledge we need to only show that for an honest pit the combined pit-reciprocal transmission is identical to that of a flat, uniform plate of no geometric structure.  This implies that the information content in the data cannot allow the inspectors to distinguish from a complex geometry (pit-reciprocal) and a flat object.

\begin{figure}[ht]
    \centering
    \includegraphics[width=0.8\textwidth]{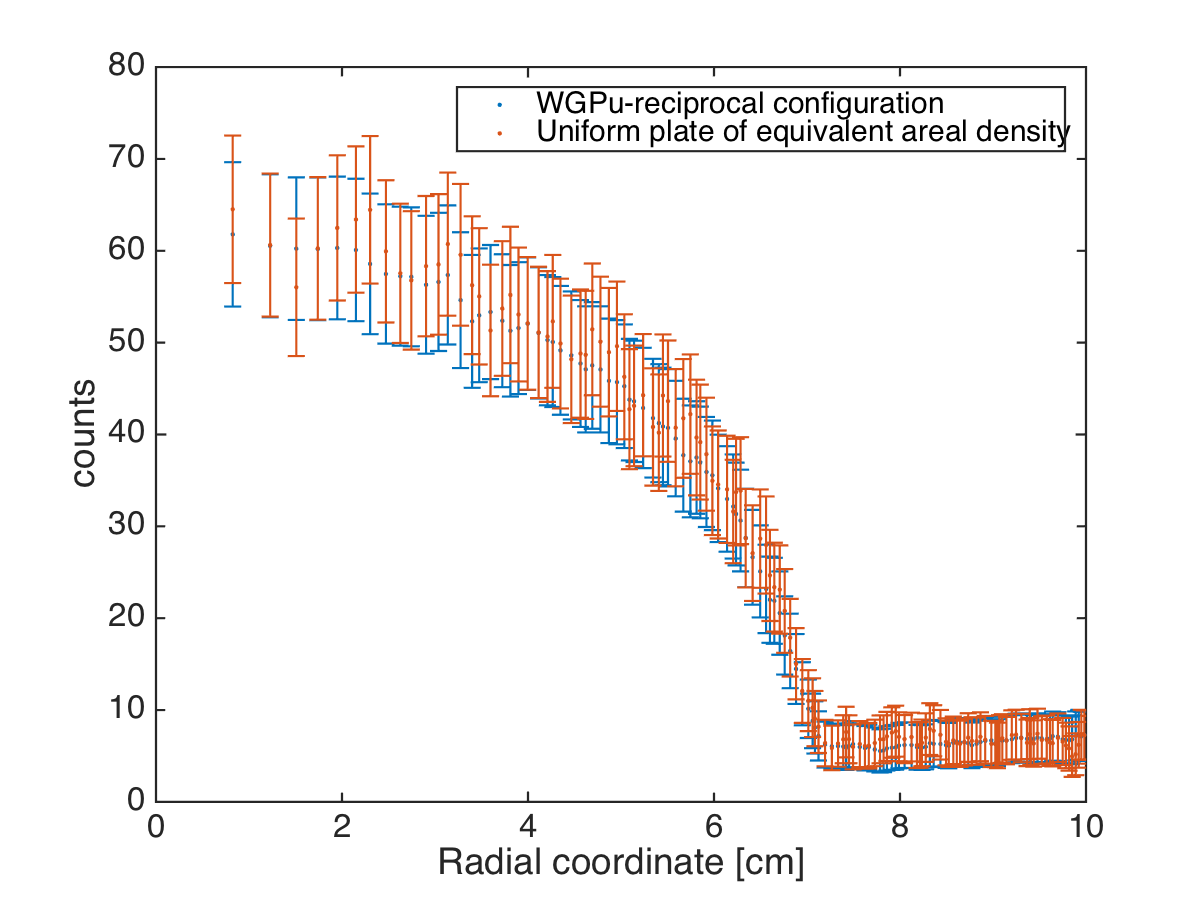}

    \noindent\caption{The radial distribution of epithermal neutron counts in the detector for the pit-reciprocal and for a flat plate of WGPu of equivalent thickness.  The error bars reflect the expected one standard deviation fluctuations for a measurement necessary for the detection of a hoax at a $5\sigma$ confidence level. The comparison shows statistically identical distributions, indicating that no information about the pit-reciprocal geometry can be extracted. }
%        \noindent\caption{Epithermal neutron images in the detector for the WGPu-reciprocal (left), for a flat plate of WGPu of equivalent thickness (right), and the radial distribution of the mean counts for both configurations(center).  The error bars reflect not the uncertainty in the MC output, but the expected one standard deviation fluctuations for a measurement with $10^5$ incident neutrons, which are necessary for the detection of a hoax at a $5\sigma$ confidence level. The comparison shows statistically identical distributions, indicating that no information about the WGPu-reciprocal geometry can be extracted.  The z-axis is the number of counts per incident neutron.}

    \label{fig:wgpu_vs_plate_2d}
\end{figure}

To show this, MCNP5 simulations have been performed for a simple flat plate of WGPu with an areal density equal to that of the pit-reciprocal configuration shown in Fig.~\ref{fig:wgpu_vs_rgpu}.  Similar to the analysis performed in the section on geometric hoax resistance the radial distribution of the counts of the two configurations are plotted in Fig.~\ref{fig:wgpu_vs_plate_2d}. The radial count distribution has error bars which reflect the fluctuations that an inspector would observe for the scenario of $1 \times 10^5$ incident neutrons necessary for achieving a hoax detection at the $5\sigma$ confidence level.  The large overlap of the errors shows the statistical identity of the two outcomes, and indicates that no useful geometric information can be extracted about the object.

\subsection*{Isotopic Information Security}
The isotopic composition of the plutonium pit can affect the reliability of the nuclear weapon~\cite{ref:explosive_properties}.  Additionally, the knowledge about the abundance of particular isotopes can allow an observer to determine the methodology by which the fissile material was produced. Because of this and other considerations the information on isotopic composition can be of sensitive nature.  Thus a zero-knowledge proof system should not produce any data from which the isotopics of the pit-reciprocal configuration can be inferred beyond what is commonly known. 

\begin{figure}[ht]
    \centering
    \includegraphics[width=0.9\textwidth]{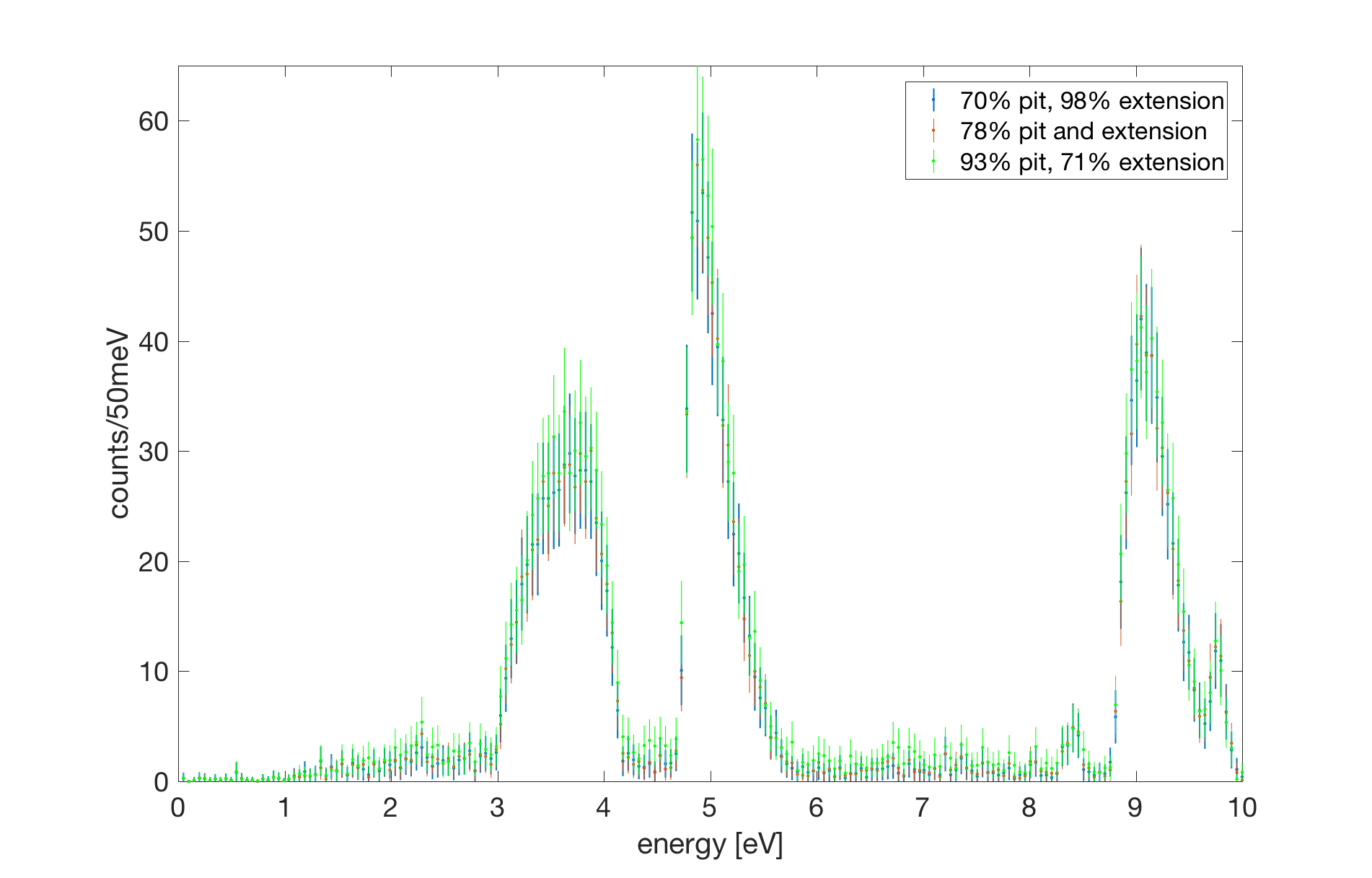}

    \noindent\caption{MC simulations of energy distribution of the transmitted epithermal neutron count mean values. The target consisted of a pit-reciprocal and an extension of various levels of $^{239}$Pu enrichment. The results show that the data would  make it impossible to determine the enrichment of the pit beyond the 70\% to 93\% range.  The error bars reflect the statistical uncertainty of an expected measurement with $1.0 \times 10^7$ incident neutrons, which is more than sufficient for achieving the 5$\sigma$ hoax detection requirement.}
    \label{fig:wgpu_plate_tamper}
\end{figure}

Just as the shape of the reciprocal mask protects the geometry of the pit, its isotopic composition can be used to mask the real isotopic concentrations in the pit itself.  While the inspectors may use the spectral information to infer the  isotopic ratios for the pit-reciprocal combination, it can be made impossible for them to infer the isotopic contributions of the pit itself.  For a simple case consider a slightly modified version of the reciprocal than the one presented in Fig.~\ref{fig:reciprocal}, where the host has added an additional flat, cylindrical extension of unknown thickness and isotopic composition.  It can be shown that multiple combinations of isotopics in the the pit-reciprocal and the extension will produce the same spectrum.  To show this computationally, MCNP5 simulations have been performed for the following three configurations:  pit-reciprocal made out of WGPu, and a 2 cm extension of 41\% enriched RGPu;  pit-reciprocal and extension at intermediate 78\% enrichment; pit-reciprocal at low-intermediate 70\% enrichment and extension made of super-grade plutonium. See supplementary material for the full listing of concentrations corresponding to different enrichment levels used in these simulations.  In the simulation incident epithermal neutron events were uniformly sampled in the $[0,10 eV]$ range.    The results of the simulations were used to determine the expected detector counts for achieving a 5$\sigma$ hoax detection.  The mean expected counts and the corresponding statistical errors are plotted in Fig.~\ref{fig:wgpu_plate_tamper}.  The plots show no statistically significant differences.  This proves that for this particular extension thickness the inspectors cannot determine the enrichment level to better than the 70-93\% range.  It can be shown that the range of uncertainty on the isotopic vector can be determined from $\Delta \mathbf{r} = \frac{y}{x} (\mathbf{r_{max}-r_{min}})$, where $x$ and $y$ are the pit-reciprocal and extension thicknesses, respectively, and $\mathbf{r_{min}}$ and $\mathbf{r_{max}}$ are vectors of the lowest and highest possible enrichment levels.  This range can be widened arbitrarily by using an extension of lower $^{239}$Pu concentration or of higher thickness.  For a more general mathematical treatment of isotopic information security, as well as a discussion on the possible use of double-chopper and velocity selector techniques for further information protection, see supplementary material.

\section*{Conclusions and Future Work}

Nuclear arms reduction treaties have long suffered from the lack of a reliable, hoax-proof and information-secure methodology for the verification of dismantlement and disposition of nuclear weapons and their components. 
The work presented here covers the basic concept behind a novel epithermal zero-knowledge verification system, which targets the fissile component of the weapon.  The Monte Carlo simulations show that epithermal neutrons can be used as a basis of a zero-knowledge proof system, allowing to authenticate a fissionable object, such as a hollow sphere,  by comparing it against another, previously authenticated template.  As required by the zero-knowledge proof, the data is physically encrypted, meaning that all the encryption happens in the physical domain {\it prior} to measurement, making it impossible for the inspection side to infer significant information about object enrichment and/or geometries.  This work has shown that the technique can be made simultaneously hoax resistant and information secure in isotopic and geometric domains.   The measurement times, using established techniques for producing pulsed epithermal neutrons, could be less than one second.

Significant additional research is needed for further understanding the strengths and limitations of this technique.  A more rigorous treatment of the geometric aspects of the proof system is necessary, to determine and quantify the resistance against possible geometric hoaxes.  This includes the task of showing that a specific measurement is unique to an object of specific isotopics and geometric shape.  In this context only hoax objects valuable from a manufacturing standpoint are of importance. The proof of uniqueness in the formalism of K-transform ~\cite{PNAS} applied to the problem of an epithermal conical beam holds promise. 

Future research should also focus on a physical proof of concept implementation of this methodology.  A source of epithermal neutrons, either using a research reactor or an accelerator-based nuclear reaction, can be used as a platform for such an implementation.  Particle detection techniques based on existing epithermal neutron detectors should also be researched and optimized~\cite{ref:losko}. The impact of of object-to-object variability on information security and specificity of this verification system should also be analyzed. 

Finally, the zero-knowledge verification technique can also be extended to weapon components made out of low-Z elements.  Most hydrogenous materials, e.g. explosive lenses, are essentially opaque to epithermal neutrons, thus necessitating the use of other, more penetrating particles.  The fast neutrons at MeV scale, where interaction cross sections for hydrogen are significantly lower, are a viable alternative for a source.  An established technique of fast neutron resonance radiography, which exploits the resonances in nitrogen, oxygen and carbon at the $\sim$MeV scale, could prove promising ~\cite{ref:blackburn2007fast}.
%%%%%%%%%%%%%%%%%%%%%%%%%%%%%%%%%%%%%%%%%%%%%%%%%%%%%%%%%%%%%%%%%%%%
%%%%%%%%%%%%%%%%%%%%%%%%%%%%%%%%%%%%%%%%%%%%%%%%%%%%%%%%%%%%%%%%%%%%
%%%%%%%%%%%%%%%%%%%%%%%%%%%%%%%%%%%%%%%%%%%%%%%%%%%%%%%%%%%%%%%%%%%%
%%%%%%%%%%%%%%%%%%%%%%%%%%%%%%%%%%%%%%%%%%%%%%%%%%%%%%%%%%%%%%%%%%%%
%%%%%%%%%%%%%%%%%%%%%%%%%%%%%%%%%%%%%%%%%%%%%%%%%%%%%%%%%%%%%%%%%%%%
%%%%%%%%%%%%%%%%%%%%%%%%%%%%%%%%%%%%%%%%%%%%%%%%%%%%%%%%%%%%%%%%%%%%
%%%%%%%%%%%%%%%%%%%%%%%%%%%%%%%%%%%%%%%%%%%%%%%%%%%%%%%%%%%%%%%%%%%%
%%%%%%%%%%%%%%%%%%%%%%%%%%%%%%%%%%%%%%%%%%%%%%%%%%%%%%%%%%%%%%%%%%%%
%%%%%%%%%%%%%%%%%%%%%%%%%%%%%%%%%%%%%%%%%%%%%%%%%%%%%%%%%%%%%%%%%%%%
%%%%%%%%%%%%%%%%%%%%%%%%%%%%%%%%%%%%%%%%%%%%%%%%%%%%%%%%%%%%%%%%%%%%
%%%%%%%%%%%%%%%%%%%%%%%%%%%%%%%%%%%%%%%%%%%%%%%%%%%%%%%%%%%%%%%%%%%%

%STOPCOUNTER

\bibliography{bibliography}

\bibliographystyle{Science}

\section*{Acknowledgments}
The authors would like to thank their colleagues at the Laboratory of Nuclear Security and Policy for their inspiration and support.  The authors are grateful for the support and encouragement from their peers within the Consortium of Verification Technologies, funded by the National Nuclear Security Administration.  The authors thank Rob Goldston from Princeton Plasma Physics Lab for valuable physics discussions and encouragement.
%This work is supported in part by the Consortium for Verification Technology under Department of Energy National Nuclear Security Administration Award DE-NA0002534.
This work is supported in part by Massachusetts Institute of Technology's Undergraduate Research Opportunities (UROP) program.

%Here you should list the contents of your Supplementary Materials -- below is an example. 
%You should include a list of Supplementary figures, Tables, and any references that appear only in the SM. 
%Note that the reference numbering continues from the main text to the SM.
% In the example below, Refs. 4-10 were cited only in the SM.    

\clearpage

\appendix
\section{Supplementary materials}
%\counterwithin{figure}{section}
%\renewcommand\thefigure{\thesection.\arabic{figure}}    
%\setcounter{figure}{0}    

\subsection*{TOF methods}

The technique described in this work requires one to determine the energy of every neutron count in the detector.  For the neutrons in the cold, thermal, and epithermal range this can be achieved via a pulsed source and time-of-flight (TOF) technique.  If the neutron pulse occurs at time $t_0$, and the non-relativistic neutron is detected at time $t=t_0 + \Delta t$ at a distance $d$ then its energy is
\begin{equation}
    E = m \frac{ (l/\Delta t)^2 }{2}
\end{equation}
where $m$ is the neutron mass and $\Delta t = t-t_0$ is TOF.  By propagating the errors we can determine the uncertainty in $E$:  
$\delta E = \delta \Delta t ml^2/\Delta t^3$.  The uncertainty in $\Delta t$ primarily comes from that of $t_0$, since most scintillation and microchannel based detectors have extremely high rise times.  Here the uncertainty on $t_0$ is either the opening time of the chopper, or the pulse length of the accelerator that produces the epithermal neutrons via nuclear reactions.  Thus, we can write 

\begin{equation}
    \frac{\delta E}{E} = 2 \frac{\delta t_0}{\Delta t}.
\end{equation}

Taking 5 eV as a point midway in our energy range, it is possible to determine the maximum time-width of the pulse to achieve the energy resolution of $\delta E =$ 0.3 eV at the distance of $l=$5 m:  $\delta t_0 = \frac{\Delta t}{2} \frac{\delta E}{E}$.  For this distance $\Delta t = 161 \mu$s, and thus $\delta t_0 = 5 \mu$s.  The precision of energy reconstruction can be increased by either making the pulse shorter, or by moving the detector further away and thus increasing $\Delta t$.    Since the geometric acceptance changes quadratically with $l$, it is statistically more optimal to shorten the pulse length than to lengthen the distance by the same fraction.

\subsubsection*{Epithermal neutron production by nuclear reactions and nuclear reactors}

The nuclear research reactors are used as very intense sources of neutrons.  Depending on the configuration, the output neutron beam's energy distribution can be thermal, epithermal, or fast.  The MIT reactor has been used to produce epithermal neutron beams for oncological applications~\cite{ref:harling}.  Using a fission plate converter, beams of $10^{10}$ n/s/$cm^2$ in the $[1 eV, 10 keV]$ range have been achieved.  The object in our study has a radius of 7 cm, thus the total flux in the $[1, 10]$ eV range will be $\sim10^9$ n/s.  This flux however will have to be modified using a chopper, in order to enable energy reconstruction via the above-described TOF techniques.  With a chopper opening of 5 $\mu s$, and a distance which corresponds to $\Delta t=161~\mu s$, the chopper will have a maximum duty of 3\%.  However, the presence of "wraparound" events, i.e. thermal neutrons with arrival times of $n \cdot 161~\mu s$, where $n$ is an integer $>1$, can introduce uncertainties in energy reconstruction from TOF. This can introduce significant backgrounds.  To avoid this, the chopper can be kept closed for $4830 \mu s$ - this will eliminate all thermal neutrons down to the energy of 6 meV, while reducing the chopper duty to 0.1\%.  Combining these numbers, the total epithermal flux of neutrons in the energy range of $[1,10]$ eV will be $10^6$ n/s.  With only $10^5$ neutrons needed for the configuration described in the main body, this translates to a measurement time of 0.1 second.

The epithermal neutrons can also be produced using nuclear reactions between accelerated light ions and various targets. There are two classes of light ion based nuclear reactions that can produce neutrons in the epithermal range.  Significant work using epithermal neutrons was performed using the Los Alamos National Laboratory's 800 MeV proton spallation source, which produces  neutrons of a broad range of energies~\cite{ref:lisowski}.  This source was the basis of a number of beam lines used for a range of applied and fundamental studies.  These included epithermal beams used for the non-destructive assay of nuclear fuels~\cite{ref:losko}.
However a particularly attractive and compact alternative to a large spallation facility are smaller proton accelerators, which allow to trigger the $^7$Li(p,n)$^7$Be and $^9$Be(p,n)$^9$B reactions.  A careful operation of incident proton energies in the initial state and neutron angles in the final state can allow to create adequate intensities of epithermal neutrons.  Ref.~\cite{ref:herrera2015new} provides calculations and data of the dependence of the double differential neutron yield on incident proton energy and emitted neutron angle. 

\begin{figure}[ht]
\centering
    \includegraphics[width=0.4\textwidth]{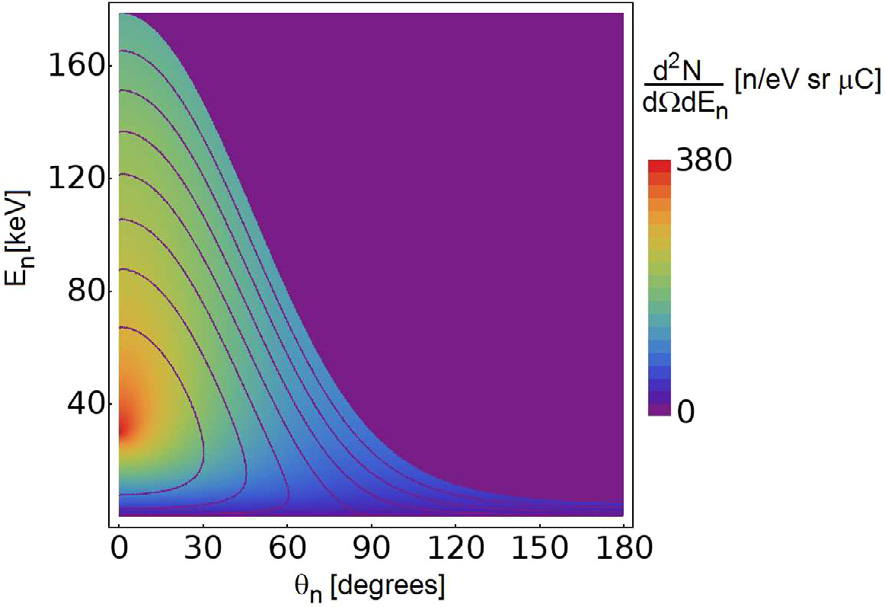}
    \includegraphics[width=0.4\textwidth]{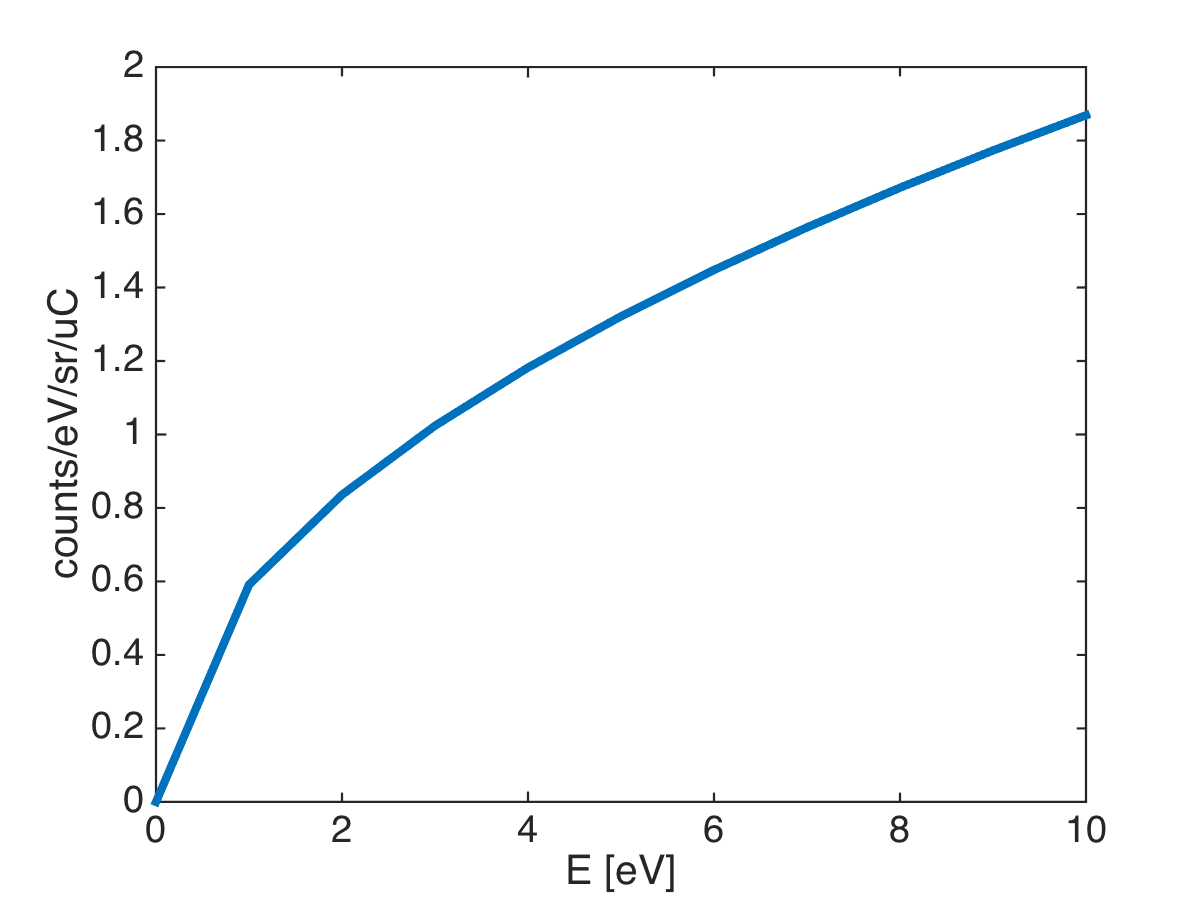}
    \caption{{\bf Left:} double differential neutron yield (colors) as function of emission angle and neutron energy for different incident proton energies on natural thick lithium target. Higher values of the incident proton beam energy broaden the accessible region. Here energies of (1.89+0.01n) MeV contours are plotted.  From Ref.~\cite{ref:herrera2015new}. {\bf Right:} epithermal neutron yields for a natural lithium target and a 2 MeV proton beam, computed for the emission angle of 90$^\circ$.}
    \label{fig:herrera}
\end{figure}

Fig.~\ref{fig:herrera} shows the double differential yield, plotted against emitted neutron energy and angle, as well as plotted against neutron energy in the $[0,10]$ eV range for the emission angle of 90$^\circ$.  Using the 1 count/eV/sr/$\mu$C as the order-of-magnitude value and assuming a beam opening of 30$^\circ$  the expected total epithermal yield in the $[0,10]$eV energy range incident upon the target will be $r=8500~s^{-1}mA^{-1}$.  Some off-the-shelf commercial accelerators can produce 2 MeV proton beam currents of  $\sim50~$mA~\cite{ref:dynamitron,ref:matsuyama}.  It was determined that about $10^5$ incident neutrons are necessary for achieving rejection of hoaxes at the 5$\sigma$ confidence level. Taking a proton beam current of $1$ mA,  this translates to a measurement time of just 12 seconds.

While the $^7$Li(p,n)$^7$Be is an attractive reaction, a number of other, similar reactions exist, such as $^9$Be(p,n)$^9$B.  Some of these may have higher epithermal neutron yields.  The search for an optimal reaction is outside of the scope of this work and may be a subject of future studies.

A significant challenge when using TOF techniques is the presence of the thermal neutrons, which arrive at times much longer than the waiting time for the epithermal pulse, thus making it difficult or impossible to identify the originating pulse.  This could introduce uncertainties in the TOF reconstruction.  However the calculations shown in Fig.~\ref{fig:herrera} show that the $(p,n)$ reactions have almost no thermal flux, thus significantly limiting their impact on the precision of the TOF method.

\subsection*{Alignment and variations in design}

Two issues come to the fore when discussing the use of reciprocals for a zero-knowledge proof system.  As in any detection system, this verification system's sensitivity will have its limits, affecting the inspector's ability to distinguish between objects of various sizes or different isotopic concentrations.  The detection probability of the system is determined by measurement times, detector sensitivities, as well as the specificity of neutron interaction physics.  These factors are mostly of stochastic nature, and measurements of arbitrary sensitivity can theoretically be achieved by varying the measurement times.  However, systematic effects are also present.  These include the unit-to-unit variability, due to manufacturing precision, as well as the hosts' ability to align the template and the candidate with the reciprocal.  Modern surveying methods allow alignment precision down to the $\mathcal{O}(10\mu m)$ scale, thus the main difficulty is related to the actual unit-to-unit variability.  

Information on manufacturing precision are not available in open domain.  For a given variability, the two sides can agree to broadened criteria of verification in order to accommodate such variability and thus avoid false alarms and reveal information about the geometry of the pit.

This circumstance in its turn limits one's ability to achieve arbitrary sensitivity.  It is reasonable to assume that manufacturing variations are small, and thus the hoaxing scenarios attainable due to this limit on sensitivity are probably not of a significant advantage to either side in the inspection regime.  A more rigorous treatment of this problem should be part of future research, possibly in the classified domain.

%\subsection*{Chopper based methods using a reactor neutron source}
%\subsection*{Combining choppers and pulsed accelerators for information security}

\subsection*{Probabilistic Tests and minimum necessary counts for 5$\sigma$ detection}

For a particular energy bin $i$, the statistical significance in units of sigma can be determined via $n_i=(c_{0,i}-c_{1,i})/\sqrt{c_{0,i}+c_{1,i}}$, assuming Poisson statistics, where $c_0$ and $c_1$ are the counts from the two distributions undergoing comparison.  In frequentist statistical analysis, and assuming normally distributed errors we can determine the probability that the disagreement between $c_{0,i}$ and $c_{1,i}$ is consistent with the null hypothesis, i.e. is caused purely by statistical fluctuations.  Conversely, the confidence for rejecting the null hypothesis and accepting the anomaly hypothesis (e.g. a hoaxing scenario is underway) can be determined via $p_i=1-C(0,n_i)$, where $C(0,x)$ is the cumulative distribution. For example, an $n=5(\sigma)$ outcome (used in high energy physics for identifying new particles) corresponds to a confidence level of $p=1-C(0,5)=1-2.9\times10^{-7}$, a very high confidence that can be used as the standard of testing.

For a multi-bin data the more common test is the chi-square test.  If the data has N bins, the number of degrees of freedom (NDF) is N.  The probability that two distributions are deviating only due to normal fluctuations can be determined from $p=Prob(\chi^2,NDF)$, where $Prob(x,y)$ is the chi-square distribution and $\chi^2$ is the (non-reduced) chi-square that can be computed from $\chi^2=\sum_i^{NDF} (c_{0,i}-c_{1,i})^2/(c_{0,i}+c_{1,i})$.

For the data presented in Fig.~\ref{fig:wgpu_vs_rgpu} we have NDF$=202$ and $\chi^2=67177$.  For this value of NDF the value of $\chi^2$ corresponding to the confidence of $1-2.9\times10^{-7}$ (the 5$\sigma$ standard) is just $\chi^2|_{5\sigma}=319$.  Clearly the discrepancy observed here implies an almost complete agreement with the anomaly hypothesis. Furthermore, it is possible to determine the minimum statistics necessary to bring the $\chi^2=67177$ result, achieved by using $N=2.0\times10^7$ neutrons, to a $5\sigma$ result.  Since $\chi^2$ depends linearly on the statistical count, the fraction of statistics necessary is just $n=N(319/67177)$, i.e. just $n=1.0\times10^5$ incident neutrons.  Most epithermal neutron sources can produce this neutron count in a matter of minutes.

\subsection*{Concentrations of Plutonium isotopes used in Isotopic Information Security Analysis}

Table~\ref{tab:concentrations} lists the concentrations of individual isotopes of plutonium for the various levels of enrichment used in the isotopic information security analysis.  For all the objects the density was 19.8 g/cc.

\begin{table}[!h]
\hspace*{-18pt}

  \centering
  \begin{tabular}{l c c c c c}
    \toprule
%    Total areal    &  \multicolumn{2}{c}{F.O.M. for Sn and W} \\
  &  $^{238}$ Pu & $^{239}$Pu & $^{240}$ Pu & $^{241}$Pu & $^{242}$Pu \\
    \midrule
Super-grade  & - & 0.98 & 0.02 & - & - \\ 
Weapons-grade (WGPu) &    0.0005    &    0.9353 &    0.0598 &    0.004 &    0.0004
 \\
Intermediate        &    0.0088 &    0.784 &    0.1304 &    0.0453 &    0.0315
\\
Low-intermediate    &    0.0124 &    0.7057 &    0.1745 &    0.0634 &    0.0441
\\
Reactor-grade (RGPu)       &    0.0297 &    0.4059 &    0.3069 &    0.1485 &    0.1089
\\

%94238 0.000502
%       94239 0.935270
%       94240 0.059767
%       94241 0.003968
%       94242 0.000494 
       
\bottomrule
  \end{tabular}
  \caption{The concentrations of plutonium isotopes for various levels of enrichment used in the paper.    The concentrations are hypothetical, and are based on Refs.~\cite{ref:explosive_properties,ref:PNNL_enrichments}.}
  \label{tab:concentrations}
\end{table}

\subsection*{Isotopic Information Security}

The loss of a beam neutron due to some form of interaction can be described by the attenuation factor $A= \frac{I}{I_0} = \exp{(-\mu \rho d)}$, where $\rho$ and $d$ are the density and thickness of a medium, $\mu = \sigma N_{A}/A$ is the mass attenuation coefficient, $N_{A}$ is Avogadro's number, $A$ is the atomic number and $\sigma$ is the total energy-dependent interaction cross section.  This is only an approximation, because it treats all elastically scattered neutrons as undetected.  For a transmission detector with a small  acceptance this can nevertheless be a good approximation for an analytical treatment of the dynamics of isotope-dependent transmission.

\begin{figure}[ht]
    \centering
    \includegraphics[width=0.9\textwidth]{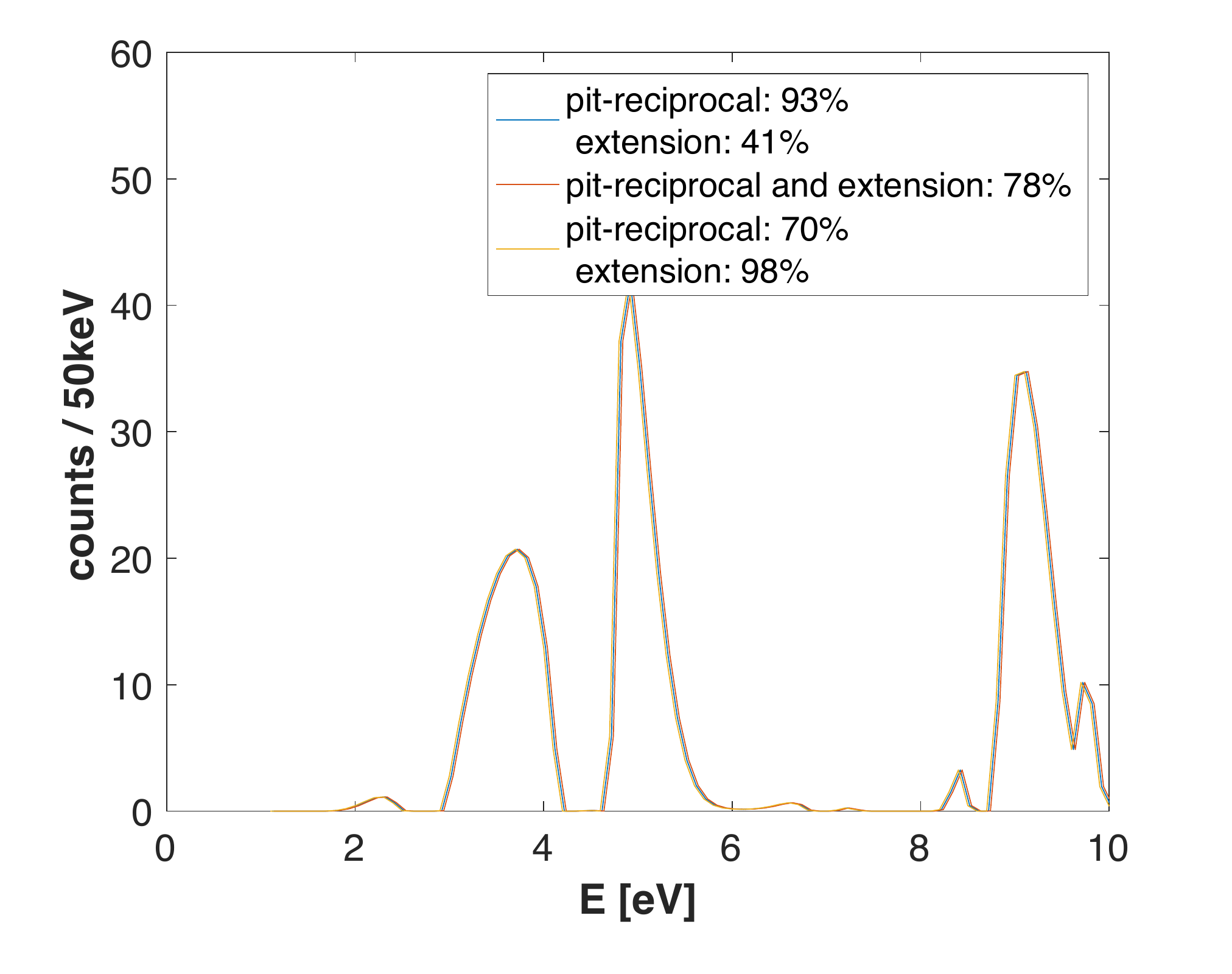}
    \caption{Transmitted calculated spectra for Scenarios 1, 2, and 3, for a total of 10mln incident epithermal neutrons uniformly sampled in the 0-10 eV energy range.}
    \label{fig:isotopic_is_calculation}
\end{figure}

Consider a particular material with isotopic vector $r_i=\{r_{238},r_{239},...,r_{242}   \}$, where individual elements are the fractional concentration of a particular isotope, such that $\sum_i r_i =1$.  Then the attenuation in a particular energy bin can be determined from 
\begin{equation}
    A(E)= \exp{\{ -\rho d \sum_i r_i \mu_i(E) \}}.
\end{equation}
If the object consists of a pit-reciprocal combination of thickness $x$ and a vector $r_i$, an extension plate (see the main body for an explanation) of thickness $y$ and a vector $r'_i$, then the logarithm of total attenuation is simply

\begin{equation}
    \ln{ A(E)}=  -\rho  \sum_i \mu_i (E)z_i(x+y)
\end{equation}
where $z_i = (x r_i   + y r'_i)/(x+y)  $ is the effective isotopic concentration vector of the combined pit-reciprocal-extension.  It can be shown that infinite combinations of $x,y,r_i,$ and $r'_i$ will produce the same value of $z_i$.  To illustrate this consider three scenarios for $r_i$, $r'_i$, and $x=5$cm and $y=2$cm:
\begin{enumerate}
\item{Scenario 1:} $r_i$ and $r'_i$ correspond to WGPu and RGPu.  See Table~\ref{tab:concentrations} for detailed values.  In this case $z_i$ corresponds to intermediate-grade plutonium.
\item{Scenario 2:} Both $r_i$ and $r'_i$ are simply equal to $z_i$ from Scenario 1, i.e. correspond to intermediate-grade plutonium.
\item{Scenario 3:} $r'_i$ corresponds to super-grade plutonium.  Using $z_i$ from the above scenarios, we find $r_i$ to correspond to low-intermediate grade plutonium.
\end{enumerate}
In  these three scenarios the enrichment levels for the pit varied between 70\% and 93\%, while the effective isotopic concentration vector remained constant at $z_i=\{0.0088,0.784,0.1304,0.0453,0.0315\}$.  Thus all these scenarios will produce the same transmission spectrum.  Fig.~\ref{fig:isotopic_is_calculation} shows the results of calculations of transmitted spectra for these three scenarios, showing identical transmitted outputs.  

%% Work on this if you have time
In addition to demonstrating this concept via calculations or MC simulations, it is also possible to determine the maximum range of uncertainty for reconstructed $r_i$, given actual values of $\mathbf{r}$, $\mathbf{r'}$, and corresponding $\mathbf{z}$.  The possible values of $\mathbf{r}$ are limited by
 \begin{align*}
    \mathbf{r_{min}}=\frac{\mathbf{z}(x+y)-\mathbf{r'_{max}}y}{x} \\
    \mathbf{r_{max}}=\frac{\mathbf{z}(x+y)-\mathbf{r'_{min}}y}{x}.
\end{align*}
Here $\mathbf{r'_{max}}$ and $\mathbf{r'_{min}}$ correspond to all {\it possible} enrichment levels from which the extension can be made.  The maximum is then just the super-grade plutonium, while the minimum can be the reactor grade plutonium. 
The full range of values of $\mathbf{r}$ is then simply
\begin{equation}
    \Delta \mathbf{r} = \mathbf{r_{max}} - \mathbf{r_{min}} = \frac{y}{x} (\mathbf{r'_{max}-r'_{min}}).
\end{equation}
By using the values of $x=5$ cm, $y=2$ cm, and solving for the $^{239}Pu$ enrichment $r_{239}$, we find that the range corresponds to about $\Delta r_{238}=23$\%, which is consistent with the previous result of 70-93\%.  This range can be further widened by either increasing $y$, or using $r'_{min}$ of even lower enrichment.  

As already stated, the simple calculation doesn't take into account such effects as in-scatter by neutrons.  This necessitates a more thorough MC simulation to fully validate this idea.  Such a simulation was performed using the MCNP5 package, and the results can be seen in Fig.~\ref{fig:wgpu_plate_tamper} in the main body.  The simulations confirm the conclusion of the analytic calculations above.

The importance of the above treatment is great:  while the inspectors can use the data to reconstruct $z_i$, they will not be able to reconstruct $r_i$ beyond simply stating that the pit enrichment level is somewhere between 70\% and 93\%.  The knowledge of this broad range is essentially useless information, as it is already known that the plutonium in most weapons is at the WGPu enrichment levels.  
This range can be further broadened, if necessary. As discussed above that can be achieved either by using an extension of lower enrichment level or one of a thicker value of $y$ - albeit at the need for longer measurement times.  Finally, the reciprocal mask itself can be made modular:  the recessed area shadowing the pit can be made of $r'_i$, while the peripheral part can be made from $z_i$ - thus removing the need for an extension plate.

While the analysis above shows that it is possible to protect the absolute isotopic information, some information about pit-to-pit variability may be inferred by the inspectors from comparative analysis of transmission spectra, for example by observing the variability in the absorption lines due to variable concentrations of $^{241}$Pu, which has strong resonances at 4.2 and 8.5 eV.  To mitigate this, the hosts and the inspectors could agree to a reduced resolution, as a way of "smearing" the absorption lines from that particular isotope.  There are a few ways of achieving this.  One approach would be to broaden the $t_0$ in the TOF technique by using a broader proton pulse for a $^7$Li(p,n)$^7$Be reaction~\cite{ref:herrera2015new}.  For the case of a chopper technique a wider slit can be used.

If necessary, the information security of the system can be further strengthened by extending the the epithermal neutron source in this proof system with a velocity selection.  Velocity selection is a well established technique for filtering out neutrons based on their energy/velocity.  A velocity selector is a system of multiple blades whose length, pitch angle and angular velocity allows only neutrons of a particular velocity range to pass through~\cite{ref:MAMONTOV}. A yet simpler configuration would consist of two choppers:  the first one setting the $t_0$, and the second one, with a phase shift, selecting the neutrons based on their arrival time and thus their energy.  Such a device could serve as a physical information barrier, allowing the hosts to limit the measurement to a particular pre-negotiated spectral region(s). Meanwhile the inspector can measure the velocity explicitly via the TOF information, as a way of confirming that the prover is not manipulating the output window of the velocity selector.

\subsection*{Reciprocal Geometries}

\begin{figure}[ht]
    \centering
    \includegraphics{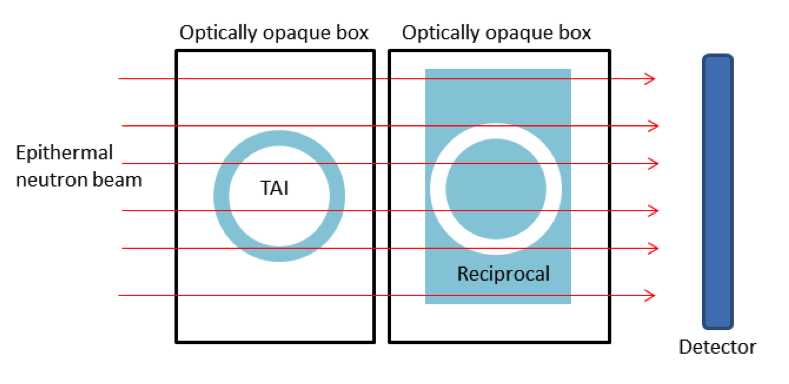}
    \caption{The verification concept employs a reciprocal mask which is designed such that the areal-density of a valid pit and reciprocal is uniform when viewed along the beam axis. When radiographed with epithermal neutrons, the resulting detector image is uniform and matches that of a plate of the same areal density.  Here TAI (treaty accountable item) refers to the pit. To maintain secrecy, both the pit and the reciprocal should remain at all times in optically-opaque boxes.}
    \label{fig:simple-reciprocal}
\end{figure}

The main function of the so-called reciprocal mask is to make it impossible for an observer to extract any sensitive isotopic or geometric information about the pit from a direct transmission measurement of the combined pit-reciprocal geometry.  The simplest way of achieving this is by taking a space encompassed by a rectangular prism,  filling it with a shape identical to the pit but with the negativer of its density, then adding a uniform density until all the negative density voxels have zero density.  This amounts to creating the negative of the pit.  The 2-d cutaway of such a simple approach can be seen in Fig.~\ref{fig:simple-reciprocal}.

While intuitively simple, this particular type of reciprocal mask has a number of problems.  For example, it would be very hard to keep subcritical.  Even if the criticality of the mask can be significantly reduced (e.g. by slicing it perpendicular to the beam axis and introducing space between the slices), the combined thickness of pit-reciprocal configuration is unnecessarily high, thus necessitating long measurement times for a statistically significant detection.

\begin{figure}[ht]
    \centering
    \includegraphics[]{p21.png}
    \includegraphics[]{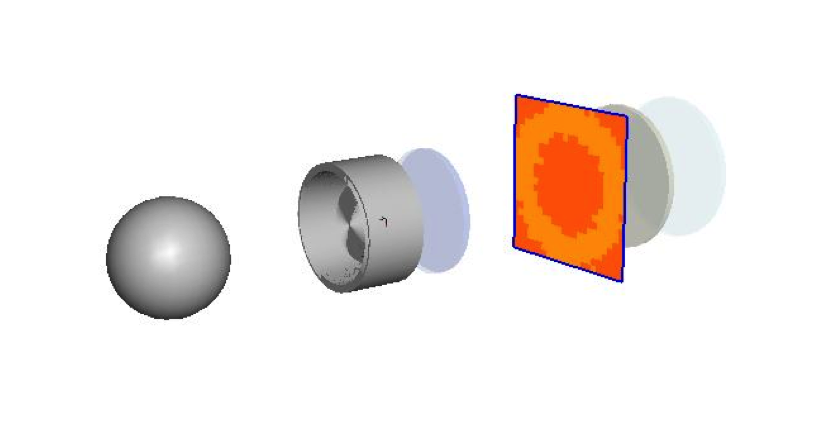}
    \noindent\caption{{\bf Top:} a diagram of the pit and its reciprocal mask, aligned along the axis of the interrogating beam.  The combined transmission image will be identical to that of a flat plate with a thickness equal to the external thickness of the mask. {\bf bottom:} the 3D view of the pit and the reciprocal. }
    \label{fig:reciprocal2}
\end{figure}

A much more optimal reciprocal mask can be built simply by realizing that the thickness of the mask along a transmission axis needs to be equal to $D-z$, where $D$ is some constant combined thickness and $z$ is the thickness of the pit along that axis. So, for a hollow shell of internal and external radii $r_1$ and $r_0$ the reciprocal can be defined via its thickness along the beam axis $d=D-2(\sqrt{r_0^2-y^2}-\sqrt{r_1^2-y^2})$ for $y<r_1$ and $d=D-2\sqrt{r_0^2-y^2}$ for $r_1 \le y \le r_0$, where  $y$ is the vertical coordinate.  A combination of the pit and the reciprocal is illustrated in Figure~\ref{fig:reciprocal2}.  For this particular case the combined thickness amounts to $D=5$~cm. 

The geometry and the enrichment of the combined pit and reciprocal is important when it comes to safety considerations.  As suggested earlier, the wrong geometry may either be too close to criticality, or simply impossible to construct.  Thus the criticality analysis of the geometries needs to be performed.  As a neutron is incident on the pit or the reciprocal, it can trigger neutron induced fission, leading to a fission chain.  The time dependence of the chain and the number of fissions can be determined from $N(t)= \exp{(k_{eff}-1)t/\tau}$, where $t$ is time, $\tau$ is the mean lifetime of a neutron in the order of 10 ns, and $k_{eff}$ is the k-effective.  Positive values of $k_{eff}-1$ cause the reaction to quickly diverge in what is called a criticality event.  For example for $k_{eff}=1.1$ it would take less than a microsecond for all nuclei in the pit to undergo fission, resulting in a nuclear explosion.  On the other hand, for values of $k_{eff}=0.9$ the chain will exponentially decay with the lifetime of $\sim$100 ns.

To determine the feasibility and the safety of the proposed configuration a set of MCNP5 simulations were performed to determine the $k_{eff}$.  For a geometry described in Fig.~\ref{fig:reciprocal2} and made of WGPu the k-effective was determined to be $k_{eff}=0.866 \pm 0.001$. A criticality analysis was also performed on the 78\% enrichment configuration described in Fig.~\ref{fig:wgpu_plate_tamper}, where a 78\% enriched pit and reciprocal are followed by a 2cm extension of the same enrichment level.  For this configuration $k_{eff}=0.8318 \pm .0002$.  For comparison, the new graphite pile at MIT's Nuclear Reactor Lab has $k_{eff} \approx 0.82$~\cite{ref:kord}.  It is not shileded, is open for general access and for educational purposes, and doesn't require any certification or regulatory oversight.  To explore ways of further reducing this number, the reciprocal geometry in Fig.~\ref{fig:reciprocal2} was modified by breaking it town into individual concentric hollow cylinders, which have been extended along the z-axis in a "telescope" configuration.  Such a modification significantly drops k-effective, bringing it to  $k_{eff}=0.621 \pm .001$.  In conclusion, the assemblies used in the concept described in this work are safe from the point of view of criticality consideration.

\clearpage

\end{document}